\documentclass[onecolumn,aps]{revtex4}
\usepackage{graphicx}
\usepackage[cmex10]{amsmath}
\usepackage{amssymb}
\usepackage{algorithm}
\usepackage{algorithmic}
\usepackage{array}
\usepackage{url}

\newcommand{\ket}[1]{|#1 \rangle}

\begin{document}

\title{Mapping of Topological Quantum Circuits to Physical Hardware}
\author{Alexandru Paler$^1$, Simon J. Devitt$^2$, Kae Nemoto$^2$ and Ilia Polian$^1$}

\affiliation{$^1$University of Passau,
Innstr. 43, 94032
Passau, Germany\\
$^2$National Institute of Informatics, 2-1-2 Hitotsubashi, Chiyoda-ku, Tokyo 101-8430, Japan.}

\begin{abstract}
Topological quantum computation is a promising technique to achieve large-scale, error-corrected computation.  Quantum hardware is used to
create a large, 3-dimensional lattice of entangled qubits while performing computation requires strategic measurement 
in accordance with a topological circuit specification.  The specification is a geometric structure that defines 
encoded information and fault-tolerant operations.  The compilation of a topological circuit is one 
important aspect of programming a quantum computer, another is the mapping of the topological circuit into the 
operations performed by the hardware.  Each qubit has to be controlled, and measurement 
results are needed to propagate encoded quantum information from input to output.  In this work, we introduce an algorithm for 
mapping an topological circuit to the operations needed by the physical hardware.  We determine the control 
commands for each qubit in the computer and the relevant measurements that are needed to 
track information as it moves through the circuit.  
\end{abstract}

\maketitle

Quantum computing is at a pivotal point in development.  The major
enabling factor is the greatly enhanced understanding of highly
efficient quantum error-correcting codes which allow reliable
operation of a quantum computer even when its individual qubits
have very large error-rate. This paper focuses
on the generic class of measurement-based topological
quantum computers \cite{DKLP02,FSG08,RHG07,FG08} (TQC), which combine several
advantages that make them one of the most promising paradigms
to build scalable quantum computers in the near future 
\cite{SJ08, DFSG08, MLFY10, JMFMKLY10, YJG10,LJLNMO10,FM13,N13, FM12}. TQC is
based on advanced topological quantum
error-correcting codes, where the quantum hardware produces 
a well defined array of qubits that are arranged and connected using 
quantum entanglement links in a regular 3D
lattice. 

TQC allows, in principle, a reliable realisation of large
quantum algorithms, such as the Shor algorithm to factor numbers
in polynomial time \cite{S94,DSMN12}, a task that is not efficiently possible on a classical
computer. However, implementing complex functionality requires
specific design flows based on clear abstraction levels and
automated synthesis steps. In case of classical circuits, the
design stack typically includes algorithmic level,
register-transfer level, gate level, transistor level, and mask
level (sometimes, more levels are defined). In over 50 years of
classical circuit design, a large variety of automated and
semi-automated methods to translate a design from a higher level
to a lower level have been developed. 

\begin{figure}[t!]
\centering
\includegraphics[scale=0.6]{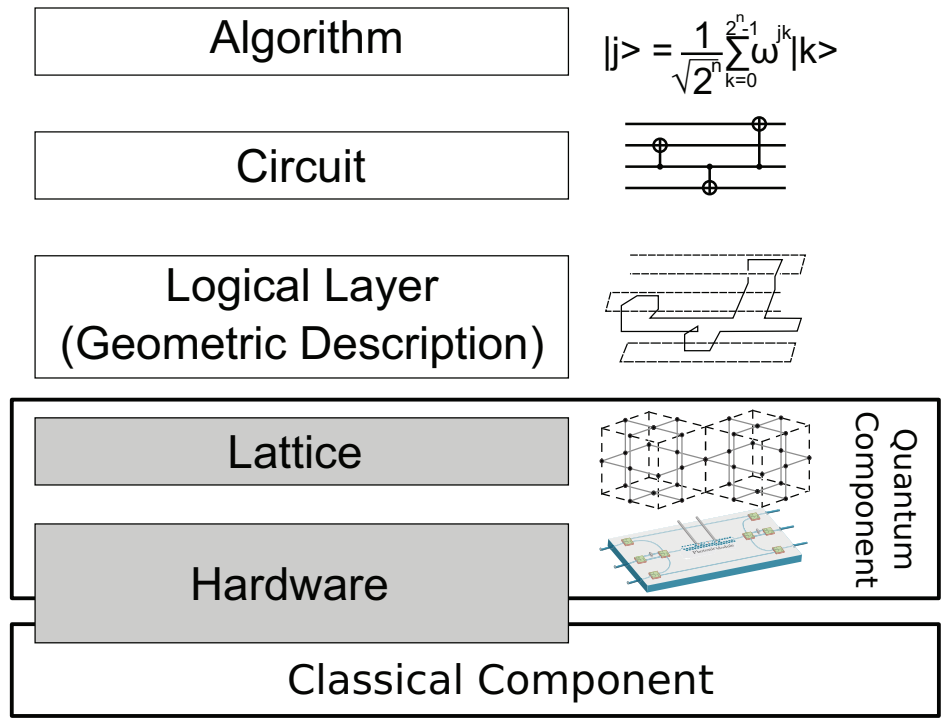}
\caption{\textbf{The TCQC design stack.} At the lattice and Hardware levels, there are two elements to the design stack; one classical and one quantum.  The 
quantum component is the instruction sets that tells each qubit in the computer how it is measured.  The classical component is keeping track of measured data in order to successfully propagate information from input to output in the topological circuit.}
\label{stack}
\end{figure}

For quantum algorithms based on the TQC paradigm,  a design
stack can also be defined (Figure \ref{stack}). A quantum algorithm,
such as the Shor algorithm, is first transformed into a quantum
circuit consisting of quantum gates (each gate is a
manipulation of typically 1, 2 or 3 qubits that can be
described by a unitary matrix). This step is required for TQC
and all other quantum architectures, and substantial body of
work on gate-level design and optimisation has been published
both in Quantum Physics and in Computer Engineering community
\cite{BM04,SBM06,WIPK07,D07,MFM08,PMH08,AMMR12}.

TQC circuits allow for a generic geometric description that abstracts from
details, in particular the underlying error correction, which is
sufficient to capture relevant algorithmic interactions.  The geometric structures 
in the TQC circuit, representing encoded qubits and encoded gates 
are embedded (via physical qubit measurements) within the physical 3-dimensional lattice. 
The model provides direct support for
only a restricted subset of quantum operations, due to symmetries 
necessary for error correction to function, and if the high level circuit
contains arbitrary gates, they must first be mapped to the
supported operations using a series of well-known steps \cite{BK05+,DN06,DSMN12}. 
After that, the circuit can
be translated into its canonical geometric representation
where each gate has a corresponding region of the 3D lattice where
it is implemented.  The first automated
optimisation method for a restricted case of two-dimensional
descriptions was presented in \cite{PDNP12}.

This paper focuses on the next synthesis step: mapping of the
geometric description to the physical level. The output of the
mapping consists of two parts:  A direct instruction set for the quantum 
hardware and a classical instruction set for data processing.  The geometric 
description of the TQC circuit specifies a layout of how logical qubits should be 
arranged within the lattice of physical qubits.  The physical size and separation of these structures is proportional 
to the error correction strength needed for computation.  TQC operates by 
performing measurements on physical qubits within the lattice.  The 
type of measurement depends on whether a given physical qubits is inside or 
outside the geometric region defining a logical qubit.  The instruction set for the quantum 
hardware consists of translating the geometric structure to a specific set of individual 
measurement instructions for the physical qubits in the lattice.  

The classical data processing information is also required for byproduct
calculation.  Within a TQC circuit, large sets of 
physical qubits form what is known as a correlation surface.  These sets 
connect the inputs of a quantum circuit to the outputs.  The overall parity of these 
correlation surfaces (when they are measured) 
tells us how the logical information must be corrected in 
order to successfully propagate from input to output. 
The classical control software of a TQC must know these 
structures in order to successfully operate.  Therefore, for a TQC circuit 
specification, the determination of the physical qubit sets that define
correlation surfaces and how they interact is non-trivial and constitutes
the by far most complex part of the mapping algorithm. 


The result of the mapping algorithm is a 3D lattice of qubits
along with instructions how to use it. This is vaguely related
to the general-purpose hardware and the customised software of a
classical microprocessor. Note that TQC is technology-independent:
any implementation where qubits can be arranged in two or three dimensions 
can be used to realise this model \cite{DFSG08,YJG10,JMFMKLY10}.
Therefore, our work closes the gap between the high-level synthesis
(which reaches up the geometric description level) and
technology-oriented physical implementation approaches. It is the
foundation for the first complete automated flow from the quantum
algorithm to the implementation.

\begin{figure}[t!]
\centering
\includegraphics[width=15cm]{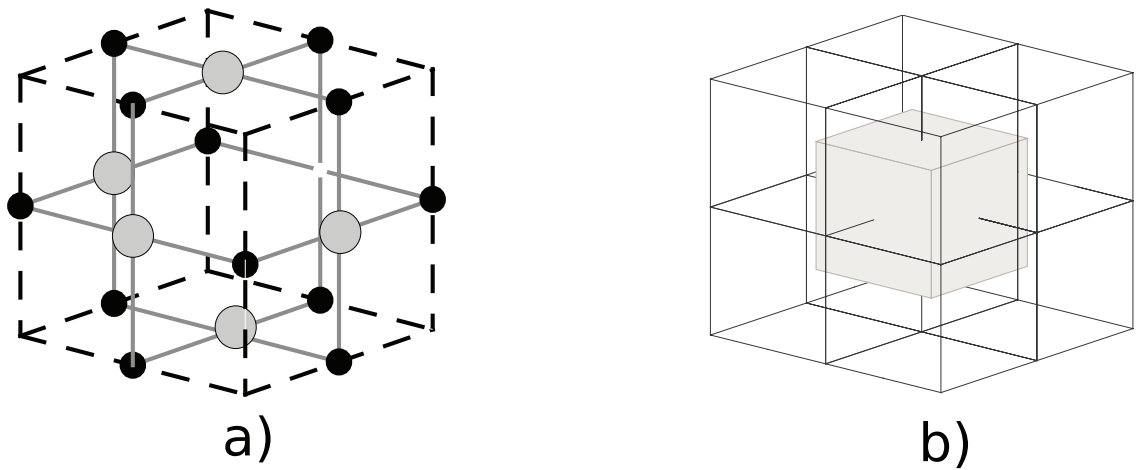}
\caption{\textbf{Structure of the Topological lattice.} a) Complete 3D lattice cell. For the cell (at lattice coordinates $cc \in CEL$) qubits marked black are $S_{cc}$, and gray qubits are $F_{cc}$; b) A primal cell (marked gray) and eight dual cells (physical qubits and entanglement are not represented);}
\label{fig:lclust}
\end{figure}

The physical qubits are arranged in a large three-dimensional lattice and
are entangled to form a single, massive, universal quantum state.  A universal 
state is a specific quantum state that can be used to implement any
fault-tolerant quantum algorithm via measurement of each qubit in the state \cite{RBB03}.  The 
unit cell of the lattice is illustrated in Fig.~\ref{fig:lclust}a). Prior to
actual computation, each qubit is initialised in the superposition
state, $\ket{+} = (\ket{0}+\ket{1})/\sqrt{2}$. Once all qubits are initialised,
a two-qubit gate is used to entangle any two qubits that are connected in
the lattice with an edge.  The unit cell is the building block of the lattice, and
its structure is repeated along three dimensions, which are denoted by
$w$ (width), $h$ (height) and $t$ (time). TQC is implemented via sequential measurement of a 2D cross
section of the full 3D lattice.  Information is propagated along the third axis of the lattice via the entanglement 
links, with processing occurring during these teleportation steps.  Hence the third dimension of the physical lattice 
is identified with the temporal axis ($t$-axis).

One general unit cell of the lattice has 27 ($3 \times 3 \times 3$) vertex positions, 18 of which are
occupied by physical qubits, denoted by gray and black circles in
Figure ~\ref{fig:lclust}a). A lattice with $mc_w \times mc_h \times mc_t$
unit cells has a total of $(2mc_w + 2) \cdot (2mc_h + 2) \cdot (2mc_t + 2)$
positions, because neighboring cells share a side. All physical qubits in the
complete lattice form the set $TQCC$. The positions not occupied by qubits
form the set $CEL$. A position $(cw, ct, ch) \in CEL$ must be the center of
a unit cell. The physical qubits at the faces
of such a cell (six gray circles in Figure~\ref{fig:lclust}a) are denoted by
$F_{cw,ch,ct}$, and the qubits at the sides (12 black circles in
Figure~\ref{fig:lclust}a) are denoted by $S_{cw,ch,ct}$. The set of all qubits
of a cell is $C_{cw,ch,ct} := F_{cw,ch,ct} \cup S_{cw,ch,ct}$. Note that
$TQCC$ is the union of these sets for all the cells in the lattice.

The topological lattice contains two self-similar lattices that are interlaced (see Figure \ref{fig:lclust}b)). The primal lattice (with corresponding $C^p,F^p,S^p$) contains all the cells having odd coordinates, and the dual lattice (with corresponding $C^d,F^d,S^d$) contains all the cells having even coordinates. If eight dual cells of the lattice are arranged into a cube, then a primal cell exists at the intersection point, and therefore the set of all physical qubit coordinates in the lattice can be expressed as $TQCC = F^p \cup F^d$.

Physical qubits (organised in cells arranged in a 3D-lattice, as described above)
are used to encode logical qubits.
The underlying lattice is a computational resource for the logical layer, similar to how the hardware is a computational resource for the physical layer. However, at the logical layer, all operations are error-corrected, implying that logical qubits are initialised, operated on and measured using logical operations. It should be noted that the logical information is not decoded prior to any logical operation and afterwards re-encoded; operations are applied directly on encoded physical qubit states.  Valid operations within the logical space are designed specifically such that physical errors do not cascade to cause uncorrectable faults.  Therefore, the TQC model 
is known as a fault-tolerant model of quantum computation.

The geometric structures, which define encoded qubits, are created by measuring 
relevant physical qubits in the $Z$ basis ($\ket{0,1}$ states).   This effectively creates a hole in the 
lattice, known as a defect.  It is the defects that encode information.  
Two types of logical qubits can be defined depending in which of the lattices the $Z$ measurements of physical qubits are performed and consequently where defects are created. The physical qubits sets $D^p \subset F^p$ and $D^d \subset F^d$ represent the physical qubits that 
are internal to defects which are measured in the $Z$ basis, with $D^p \cap S^p = \emptyset$ and $D^d \cap S^d = \emptyset$.  Primal defects 
are created by measuring qubits in the set $D^p$ and Dual defects are created by measuring qubits in the set $D^d$.

\begin{figure}[ht!]
\centering
\includegraphics[width=15cm]{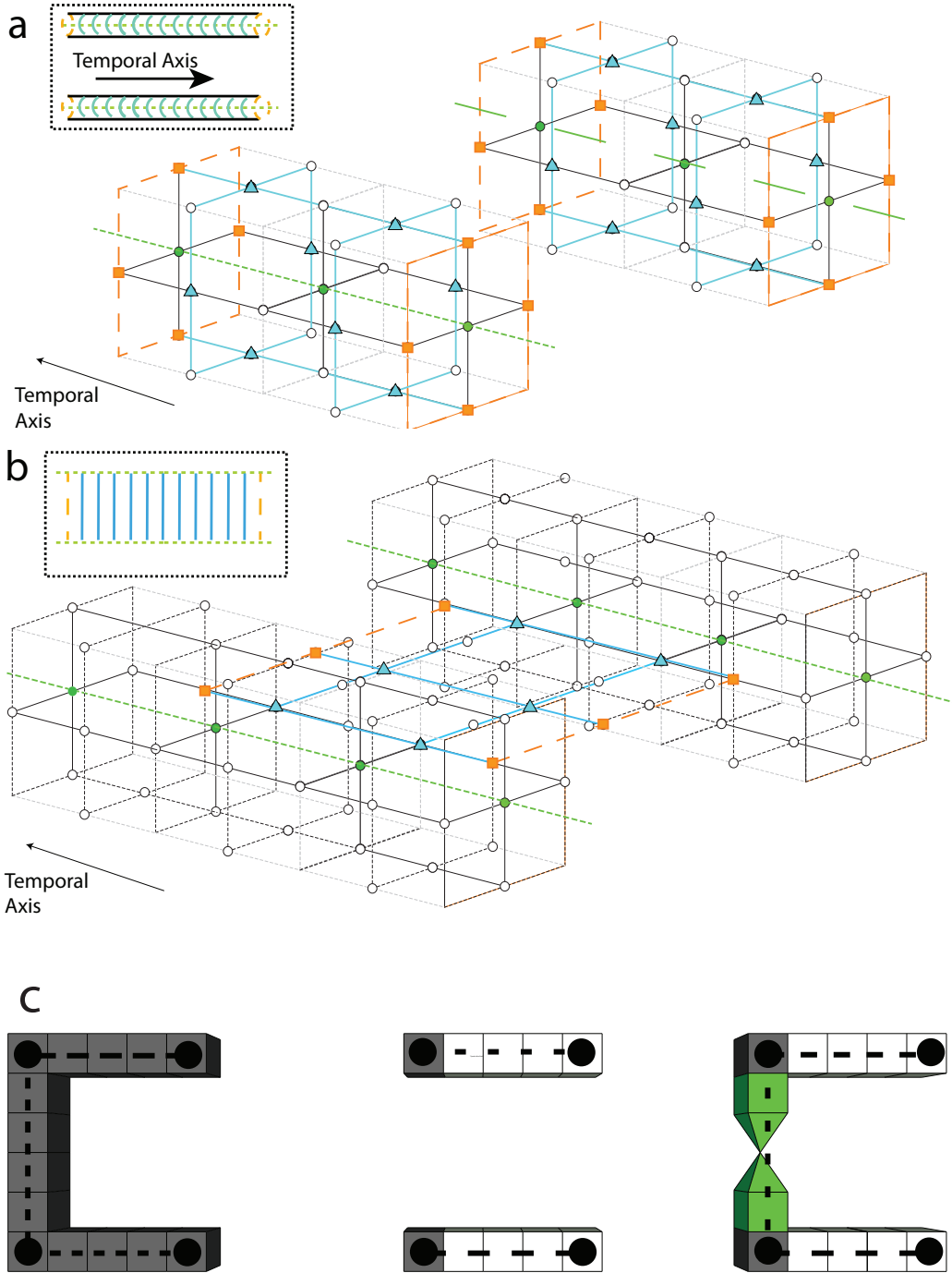}
\caption{\textbf{Qubit sets used for information processing.} Figure {\bf a)} is for loop operators (The $X$ ($Z$)-eigenstate information for primal (dual) defects) and Figure {\bf b)} is for chain operators (The $Z$ ($X$)-eigenstate information for primal (dual) defects).  
These sets define defects (green circles), logical operators (orange squares)
and correlation surfaces (blue triangles) for a single logical qubit undergoing the identity operation.  A shorthand notation is illustrated in the inserts.  
This figure illustrates for minimum sized defects (and hence the smallest distance quantum code).  Figure \textbf{c)} illustrates initialisation 
in the $Z$, $X$ basis and state injection.  State injection is where in individual qubit is measured in a rotated $Z$ basis (the vertex of the pyramids) and 
then expanded to form an encoded defect, this allows us to achieve universality.  The superimposed black vertices and lines are used to illustrate 
graph edges and vertices in the mapping algorithm. }
\label{fig:holes}
\end{figure}

Figure~\ref{fig:holes} illustrates the structure of two defects which encode a single logical 
qubit (the unit cells that are not relevant for the defects are not shown).  In this structure 
a temporal axis is defined running from the front of the structure to the rear and this 
encoded qubit is undergoing a simple identity operation. i.e. the input of this topological circuit 
starts on the 2D layer at the front of the image and via measurement is teleported to the 2D layer at the rear. 
There are three sets of physical qubits that are relevant for the mapping.  One is 
the sets $D^{p,(d)} \subset F^{p,(d)}$ that are used to define primal (dual) defects.  
These qubits are illustrated 
in Figure \ref{fig:holes} as green circles and are measured in the $\ket{0,1}$ basis to 
simply remove them from the lattice.  The second set are known as logic operators.  Logic operators are sets of physical qubits whose parity determine the logic state of the 
encoded information.  Two types of logic operators exist, one is a
loop operator that completely encircle an encoded defect.  
The second is a chain that connects both defects 
(illustrated by orange squares in Figure \ref{fig:holes}b)).

For primal (dual) defects, the loop operators correspond to the logical $X$ ($Z$) operator while the chain corresponds to the logical $Z$ ($X$) operator. The final set of relevant qubits are illustrated as blue triangles in Figure~\ref{fig:holes}. These sets are known as correlation surfaces. The correlation surface connects logic operators from an input layer in the cluster to an output layer. The ring and chain operators need to be connected from the input layer (front of the image) to the output (rear of the image). The qubits illustrated in blue perform this operation. If each qubit in blue is measured in the $\ket{+,-}$ state ($X$ basis) we connect the chain (ring) operators from the input side to the chain (ring) operators on the output side. If we wished to measure the logical state, by measuring and calculating the parity of the logical operator we would measure each physical qubit in the $Z$ basis. By measuring these qubits in the $X$ basis instead, the logical information is teleported to the next layer in the lattice, with a correction that is determined by the parity of the $X$ basis measurements. The correlation surfaces form either sheets or tubes. For primal (dual) defects, sheets are the logical $Z^p$ ($X^d$) surfaces and tubes are the logical $X^p$ ($Z^d$) surfaces. In order to track information as it moves through the lattice, we need to know which qubits are associated with these surfaces, as the parity of these measurements will dictate the correction that needs to be applied to encoded information as it moves through the lattice.

The primary goal of this work is, given a geometric definition of a TQC circuit,
to determine the information needed by the (classical) control software of a
quantum computer in order to operate TQC
hardware that implements a circuit. This information includes the measurement
basis of the physical qubits. In Figure~\ref{fig:holes}, the solid circle qubits (defect-internal)
are measured in $Z$ basis whereas the triangle qubits (correlation surfaces), square qubits
(input/output) and white qubits (qubits for error correction) are measured in $X$ basis. Moreover,
the measurement outcomes for qubits in correlation surfaces are needed 
to calculate corrections to encoded data as they propagate through more 
complicated topological structures that are not considered here.
Therefore, the mapping procedure must identify the physical qubits for each
correlation surface ($X^{p,d} \subset F^{p}$ and $Z^{p,d} \subset S^{p}$).

Other geometric structures, corresponding to specific logic operations, are shown in 
Figure \ref{fig:holes}c).  For a primal pair of defects, the horseshoe structure corresponds 
to initialising a qubit in the $\ket{0}$ state, the pair of parallel defects correspond to 
initialisation in the $\ket{+}$ state and the third structure corresponds to state injection.  
State injection allows us to introduce an arbitrary state into the TQC lattice by 
measuring a single physical qubit that exists at the vertices of the pyramid 
structure. State injection is instrumental in the TQC model 
to achieve universal computation \cite{BK05+,RHG07}.  In Figure \ref{fig:holes}c) the temporal axis runs from 
left to right.  In order to perform logical measurement, we use the exactly same structures, but in 
the reverse temporal direction (this allows for measurement in the $\ket{0,1}$ and 
$\ket{+,-}$ basis, i.e. $Z$ and $X$ bases).  The regions that are marked as grey and green correspond to the 
sets of physical qubits whose parity defines the parity of the initial encoded qubits. 

The concept of state injection in the topological model is instrumental for achieving universality.  The power of quantum 
computation is intricately related to the introduction of these injected states and their consumption to realise non-Clifford 
gates through teleportation.  However, in the context of this mapping algorithm, there is essentially no difference in the 
classical computation associated with the geometric structures in Fig. \ref{fig:holes}c).  While the right most structure in Fig. \ref{fig:holes}c) 
represents the injection of a non-trivial encoded state, the sets of physical qubits that define byproduct operators for this newly created state are no 
more complex to calculate than the sets associated with standard Clifford initialisation.  
Therefore, the introduction of injected logical states and the realisation of non-Clifford gates in a topological circuit poses no complexity problem for the 
mapping algorithm.

\begin{figure}[ht!]
\centering
\includegraphics[width=15cm]{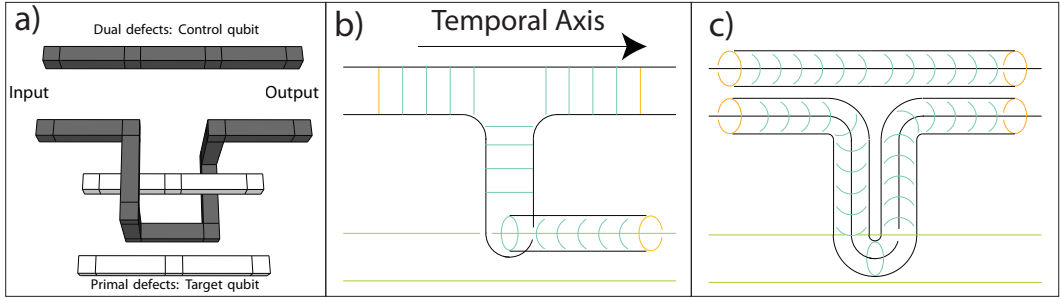}
\caption{\textbf{Evolution of correlation surfaces during a braided CNOT gate.}  {\bf a)} The geometry of a CNOT gate where the 
dark (dual) qubit acts as a control.  {\bf b)}  $X^d$ correlation surface on the control (black lines) is intersected by the target defect (green line) creating a 
correlation surface, $X^p$ in accordance with the properties of the CNOT.   {\bf c)} illustrated that the $Z^d$ surface of the control does not become correlated, as required.}
\label{fig:cnot1}
\end{figure}

The mapping problem of a TQC circuit is the accurate determination of physical 
qubits in the topological lattice corresponding to relevant correlation surfaces that 
define the operation of the circuit.  The input to this algorithm is a suitably optimised topological circuit that is compatible with the error-corrected model and hence the underlying computational hardware.  Algorithmic constraints at both the physical and logical level have already been taken into account via the construction and optimisation of the input circuit.  Before formally introducing the problem, we
illustrate it through the example of a \textsc{cnot} operation implemented via a
braid operation between two encoded defects (Figure~\ref{fig:cnot1}). A
correlation surface that begins on primal defects cannot terminate on dual
defects and vice versa. Therefore, as defects are braided, a correlation surface
extends around the qubit involved in the braid. It is this interaction that entangles
the logical qubits.

Figure~\ref{fig:cnot1} illustrates how correlation surfaces are defined, using the shorthand 
notation that is illustrated in the inserts of Figures \ref{fig:holes}a) and b).  In Figure~\ref{fig:cnot1}b) 
we show the $X^d$ correlation surface (sheet) on the control qubit (a dual type defect).  
Due to a braid with a primal defect, 
this correlation surface must extrude around the second defect and forms 
a tube.  This correlates the outputs of the two qubits when the control qubit is in an $X$-basis eigenstate.  Figure~\ref{fig:cnot1}c) shows the 
$Z^d$ correlation surface (tube) on the control qubit.  Here, the surface is 
not perturbed by the braid.  Therefore the output remains uncorrelated when the control is in a $Z$-basis eigenstate.  This circuit 
correlates output qubits in precisely the way needed to realise a CNOT 
operation \cite{NC00,FG08}.

These surfaces are not specified with the TQC circuit, and therefore the mapping
procedure must derive them
from the geometric structure and map them to the actual set of qubits that are used in the 
lattice such that they can be tracked as the circuit is implemented.

\section*{Results}
\label{sec:background}
\textbf{Problem Formalism.} The mapping procedure takes a geometric description $\mathcal{G}$ of a TQC
computation as input and calculates a data structure $\mathcal{Q}$ which contains
the required information of each logical qubit. Each logical qubit in $\mathcal{G}$ 
is specified by a set, $\sigma$, of directed segments, given by their
end points $(begin, end) \in CEL^2$. Recall that $CEL$ is the set of lattice coordinates
that are not occupied by qubits and can be centres of unit cells. Each segment has
one of six possible directions $\{\pm w, \pm h, \pm t\}$. To convert $\mathcal{G}$
to $\mathcal{Q}$, an intermediate representation $\mathcal{Y}$ where each logical
qubit is described by a directed, cyclical graph, is calculated first and then used to
derive $\mathcal{Q}$.



\textit{Geometric Description and Representation by Graphs:}
For each logical qubit $\sigma$, a directed cyclical graph $G_\sigma=(K_\sigma,V_\sigma)$
is defined. The set $K$ of possible graph vertices contains representations of unit cells from
set $CEL$. There is a bijective mapping $coord$ between $K$ and $CEL$: $coord(k)$ yields
the three ($w/h/t$) lattice coordinates of the cell corresponding to $k$. $K_\sigma \subset K$
represents the end points of the segments in $\sigma$, and $V_\sigma$ represents the connections
between the end points. A graphical example for this construction is offered in Figure \ref{fig:holes}c).

\textbf{Mapping Algorithm.}  Representation of Logical Qubits by Set Tuples:
The tuple representation $q_\sigma=(l, D^l_\sigma, I^l_\sigma, O^l_\sigma,
J^l_\sigma, X^l_\sigma, Z^l_\sigma)$ of a logical qubit $\sigma$ contains all
the information which is needed to implement it and its interactions with
other qubits on a quantum computer. The ultimate outcome of the mapping
procedure described next is the set $\mathcal{Q}$ which contains tuples
$q_\sigma$ for all logical qubits $\sigma$.


The specifics of the mapping algorithm can be found in the Methods section.  The edges in $V_\sigma$ form a Hamiltonian path, containing all vertices from $K_\sigma$. During the mapping, the path can be either modified by $remove(a); a \in K$ that removes vertex $k$ and its incident edges, and creates a new edge between its neighbours, or the $insert(a,b,c);a,b,c\in K$ function that deletes the $(a,c)$ edge, and creates the new edges $(a,b)$ and $(b,c)$.

Within $q_\sigma$, $l$ stands for the type of the logical qubit (primal or dual).
$D^l_\sigma$ includes all defect-internal physical qubits (i.e., those to be measured
in the $Z$ basis, whereas all other qubits will be measured in the $X$ basis).
$I^l_\sigma$ and $O^l_\sigma$ are physical qubits that define inputs and outputs,
respectively. In Fig.~\ref{fig:holes}, the qubits in $D^l_\sigma$ were shown in green
and the qubits in $I^l_\sigma$ and $O^l_\sigma$ in orange. Sets $X^l_\sigma$ and
$Z^l_\sigma$ include all physical qubits that are part of the $X$ and $Z$ correlation
surface, respectively. Finally, $J^l_\sigma$ is the set of injection points. Note that the
primal (dual) qubits are defined using coordinates on primal (dual) cells.


As mentioned earlier, the $Z$ correlation surface is a sheet for a primal logical qubit
and a tube for a dual logical qubit. The $X$ correlation surface is a tube for a primal
logical qubit and a sheet for a dual logical qubit. For convenience, we introduce
functions $sheet: \mathcal{Q}\rightarrow TQCC$ which returns $Z^p_\sigma$ for
primal logical qubits and $X^d_\sigma$ for dual logical qubits and $tube: \mathcal{Q}
\rightarrow TQCC$ which returns the corresponding tube in a similar way.



\begin{figure}[t!]
\centering
\includegraphics[height=6cm]{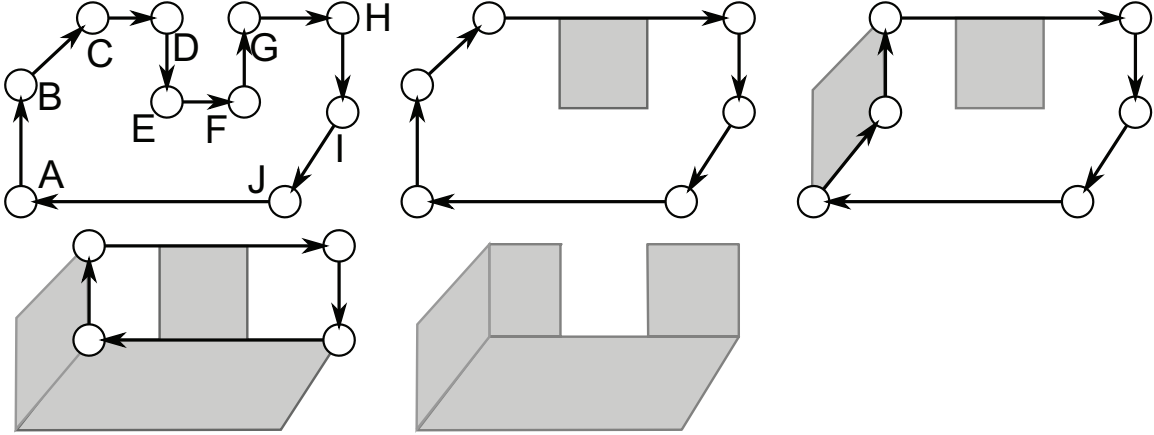}
\caption{\textbf{Example of the algorithm illustrated in Eqn. \ref{alg:sheet1}.} The above illustrates for the graph $G_\sigma=(\{A,\ldots,J\}, \{(A,B),(B,C),\ldots (J,A)\})$. For example, starting from the vertex $start=A$, the first possible operation is $reduce(E, F)$ and a first sub-sheet is found $SUBS=\{(E,G)\}$. Furthermore, because $C,D,G,H$ are co-linear will result in $remove(D)$ and $remove(G)$. The cycle is traversed until the $start$ is reached again, and the first traversal completes. After the second traversal neither $reduce$ nor $remove$ were applied. Therefore, the $reshape(A,B,C)$ is applied, and a second sub-sheet is found $SUBS=\{(A,C), (E,G)\}$. Finally the last two sub-sheets are inferred leading to $SUBS=\{(A,I), (C,I), (A,C), (E,G)\}$. The complete sheet is found by combining all the sub-sheets according to the algorithm in Eqn. \ref{alg:sheet2}.}
\label{fig:example1}
\end{figure}
\begin{figure*}[t!]
\centering
\includegraphics[width=16cm]{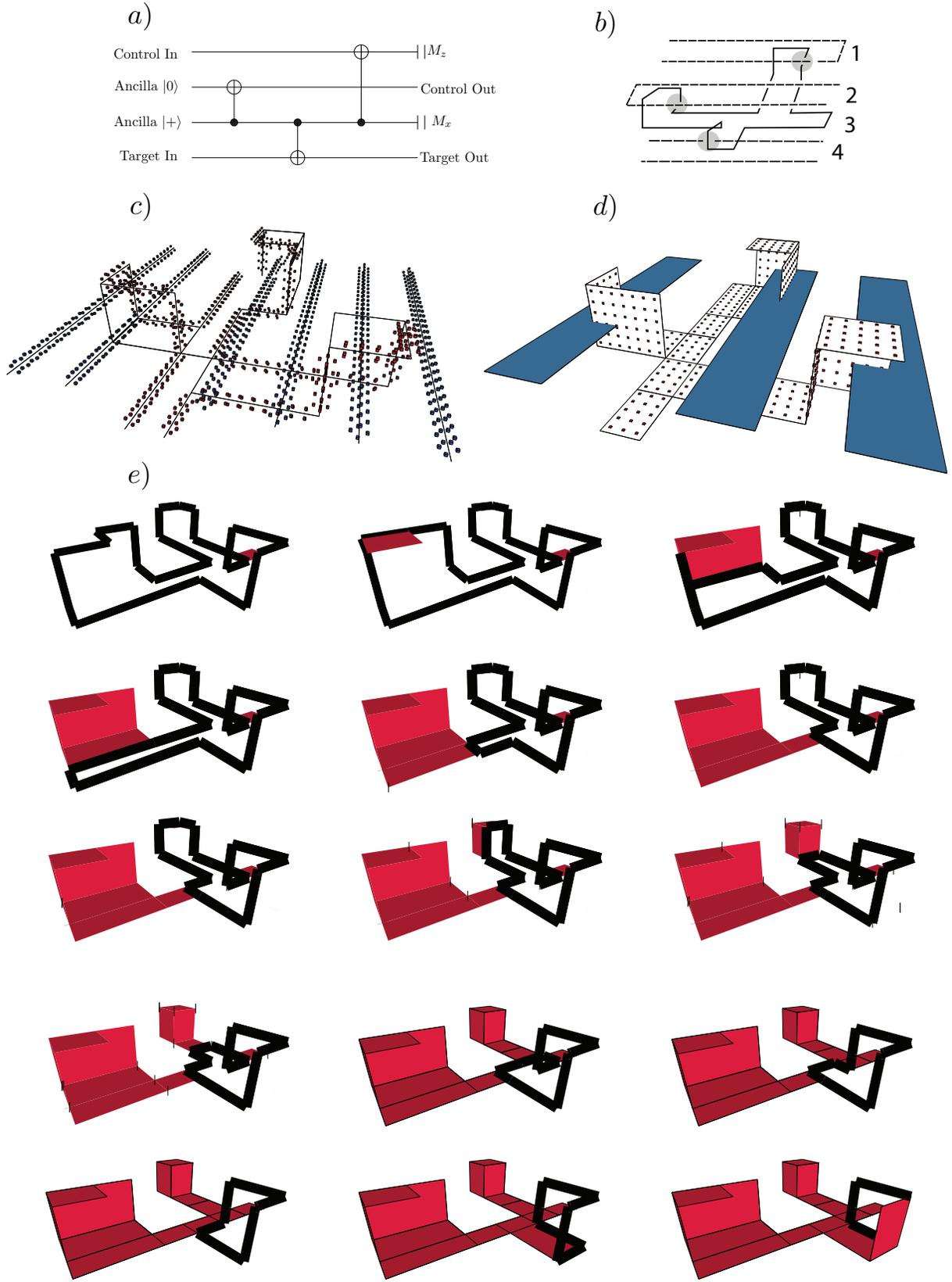}
\caption{\textbf{Defining tubes and sheets.} \textbf{a)} A standard circuit identity to perform a CNOT between two primal defects \cite{RHG07}, mediated by a single dual defect that acts as control for the three braiding operations. \textbf{b)} The geometric description of the circuit employed. The gray areas indicate the braiding of the logical qubits; \textbf{c)} The sets of qubits forming tube correlation surfaces for the topological structure: red qubits form the set of the dual surfaces, while the primal tube surfaces are formed by the blue qubits; \textbf{d)} The sets of qubits forming sheet correlation surfaces: the surface of the dual logical qubit is represented as the set of red physical qubits, while the sheets for the primal logical qubits are only illustrated. 
\textbf{e)} Intermediate steps in 
finding the sheet correlation surface for the dual logical qubit with the algorithm detailed in the methods section.  The edges of the graph used 
for computing the surface are indicated with thick lines.  This figure is slightly rotated from Figure c).}
\label{fig:tubesheet}
\end{figure*}

\textbf{Mapping Examples.} The presented algorithms were implemented and their execution will be illustrated by examples. The example from Figure \ref{fig:example1} illustrates how a complete sheet is computed. A more complex example, for the computation of tubes and sheets, is presented in Figure \ref{fig:tubesheet}, and procedure starts from the TQC computation described by the geometry presented in subfigure a). The circuit consists of 3 logical CNOTs performed between 4 logical qubits: 3 primals (the qubits $1,2,4$) and a dual (qubit $4$). The circuit will implement the following sequence: $CNOT(3,2)$, $CNOT(3,4)$, $CNOT(3,1)$.

The quantum circuit in Figure \ref{fig:tubesheet}a) is converted to a geometric form appropriate for the TQC model (Figure \ref{fig:tubesheet}b)).  The mapping will first determine the tube surfaces of each logical qubit and the results shown in Figure \ref{fig:tubesheet}c) will be obtained. Figure \ref{fig:tubesheet}d) contains the result of the physical qubit sets forming sheet surfaces. For the same dual logical qubit (the third qubit in Figure \ref{fig:tubesheet}a)) all the intermediate steps of the sheet computation are graphically presented in Figure \ref{fig:tubesheet}e) .


\textbf{Algorithmic Complexity:} The algorithmic complexity of the mapping procedure is analysed in the following. For the computation of all the qubit sets, 
computed by the algorithm in Eqn. \ref{alg:tubes}, the mapping procedure requires a single 
qubit-cycle traversal, while the computation of the coordinates is straightforward (see the illustration presented in 
Figure \ref{fig:holes}a and the corresponding legend). Therefore, for a given logical qubit $q_\sigma$,
the runtime complexity is linear in the number of vertices ($|K_\sigma|$) of the graph-cycle representing the geometry.

\begin{figure}[t!]
\centering
\includegraphics[width=15cm]{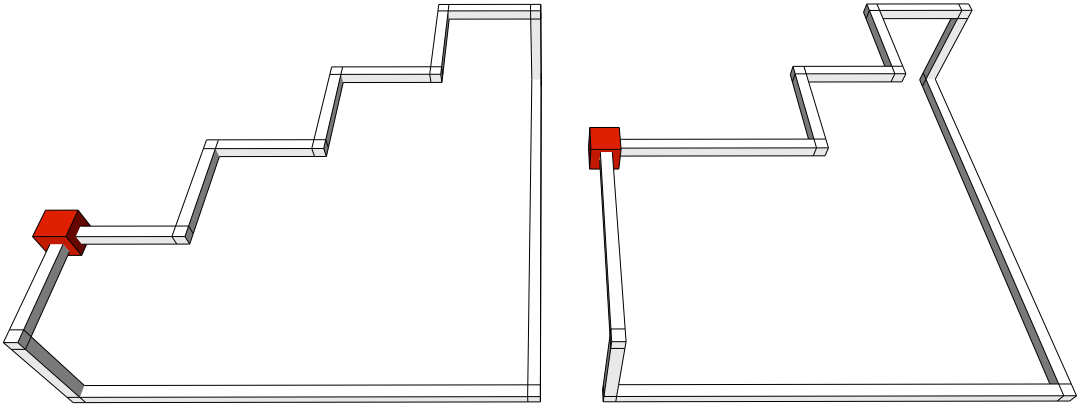}
\caption{\textbf{Worse-case geometric structure for algorithmic efficiency}.  The above is an example of defect structures that require a 
maximum number of reshape-rule applications to calculate the corresponding sub-sheets. }
\label{fig:proof}
\end{figure}

For the runtime complexity analysis of the sheet mapping procedure a \emph{worst-case geometry} has to be defined.
Such a geometry, when mapped in the 3D, will have to require a maximum number of \emph{reshape}-rule applications in order
to be able to compute the corresponding sub-sheets. The search of a geometry will consider that the algorithm from Eqn. \ref{alg:sheet1} 
randomly selects a vertex from the cycle-graph (Line 1), and that the vertex is used as a pivot for the reshape-operation. 
These assumptions imply, that even after a sequence of $reshape$ operations is applied, there is no possibility to find a sub-sheet
and to reduce the number of vertices, and thus a further $reshape$ is necessary. For a graph with $|K_\sigma|$ vertices, a worst-case
situation arises when $|K_\sigma|-3$ vertices are arranged in a pattern similar to the one in Figure \ref{fig:proof}, where the red cube indicates the
 $start$ vertex and the cycle is traversed clockwise. The graph vertices (where edges are joined) are not represented.

After a first traversal of the graph, the reduce operation could not have been applied, and a $reshape$ followed by corresponding $remove$ 
operations will transform the graph like in the right panel of Figure \ref{fig:proof}. It can be noticed, that until the step-like construction
is not reshaped, no $reduce$ can be applied. However, after each $reshape$, co-linear vertices can be $removed$, 
thus minimising $|K_\sigma|$. Overall, the complexity of the sheet-finding procedure is bounded by $O(|K_\sigma|^2)$.

While the mapping algorithm itself scales polynomially with the size of the cycle graph, we cannot make any claims regarding the size of the 
cycle graph given an optimised topological circuit.  The size of the cycle graph is essentially related to the number of 90$^\circ$ "pivots" made 
by defects in the topological circuit we do not know yet how big, in general, this set is for a given circuit specification.  Current methods in 
systematic optimisation for topological circuits are still in their infancy \cite{PF12} and determining the scaling of the number of these pivots for a given 
number of logical qubits is unknown.  However, the physical 
size of the topological circuit (in terms of the physical volume of cluster needed to implement the circuit at a fixed error correction strength) has 
marginal influence on the scaling of the mapping algorithm itself.  Instead, the scaling of the mapping algorithm will be dominated by the total number of pivots made by the input circuit specification.  

\textbf{Correctness:}  Checking the correctness of the algorithms, implies verifying their termination and the fact that the correct physical qubit coordinates are computed.
Again, for the algorithm in Eqn. \ref{alg:tubes}, the termination is guaranteed because a single traversal of the cycle-graph is necessary. 
The correctness of the computed coordinates is shown by comparing the algorithm instructions with Figure \ref{fig:holes}. The direction $d$ (Line 4)
is associated with the green line in the figure, where $b$ and $e$ are the $CEL$ coordinates of the edge vertices. For example, let us
consider that the two cells from the lower defect are $b$ and $e$. This implies, that by selecting the two neighbouring coordinates (Line 7),
the co-ordinates of the green marked qubits (the defect set $D$) is computed. The remaining $4$ qubits (Line 8) that do not belong to $D$ are the
light blue marked qubits, which are associated with the \emph{tube} correlation surface that surrounds the defect region. One has to note,
that for defects with a cross-section larger than a cluster-cell, the set union operations on Line 10 and Line 12 are to be interpreted as 
$(A\setminus\{a\})\cup((\{a\}\setminus A)\cap\{a\})$, meaning that if element $a$ existed in the set $A$ it will be removed, otherwise it will be included.

The termination of the sheet-finding algorithm (Eqn. \ref{alg:sheet1}) can be illustrated by starting from the fact that the geometric description is
mapped into a 3D representation where only 6 segment directions are possible (see the discussion of the $reshape$ and $reduce$ rules). 
A geometrically described defect configuration of a logical qubit (in the absence of any possibility to apply $reduce$ or $remove$) will 
have an even number of edges in its associated graph. Considering all the worst-case geometries (an example is Figure \ref{fig:proof}), it can be seen
that after a graph-traversal, either $reshape$ or $reduce$ (followed by $remove$) can be be applied. The number of maximum consecutive
$reshapes$ is bounded by $|K_\sigma|$. Thus, the number of graph-vertices is continuously reduced, and the termination of the algorithm is 
guaranteed.

The correctness of the coordinates computed from the sub-sheets is shown by comparing the Lines 3 and 4 of Eqn. \label{alg:sheet2} with Figure \ref{fig:holes}b). For 
a logical qubit $q_\sigma$ of type $l$, the set of all possible \emph{side qubits} (depicted gray in Figure \ref{fig:lclust} for a single cell) is $S^l$.
For the two defects from Figure \ref{fig:holes}b) the instruction on Line 3 will return the coordinates of both the orange and the blue marked qubits, 
which intersected with $S^l$ will return only the blue qubits. These are the ones necessary for the computation of a sub-sheet. If two 
sub-sheets overlap (which is possible, given the way the algorithm in Eqn. \label{alg:sheet1} functions), then the intersection set is 
not part of the set of qubits defining the complete sheet of a logical qubit (Line 4).

A final detail that needs explanation is that the approach, by finding sub-sheets, is correct. Initially, one has
to consider that for TQC, by deforming the defect geometries, does not change the computation, because the topology of the description is 
relevant, and not its geometry. For this reason, for each individual logical qubit described as a graph, the $reshape$ operation is valid,
because the sheets of two topologically equivalent geometries will differ only in sub-sheet. For example, in the last panel of Figure \ref{fig:corect}b), the 
subsheet $BD$ is the difference between the equivalent geometries: before and after applying the $reshape$ operation.

Another fact about the correctness of sub-sheet-finding appears when one considers the example presented in Figure \ref{fig:corect}a). Let us consider, the
three qubits $A,B,C \in S^l$. Deforming the defect implies, that the qubits $U,V$ are not part of the defect anymore, and that $D,E \in S^l$
 are parts of the new \emph{boundary}. This is equivalent to stating that the sub-sheet containing the qubits $D,E$ was created, and that after
the change the complete sheet will contain $(\{A,B,C\} \cup \{D,E\}) \subset S^l$.

It should be noted that the correctness of the mapping algorithm already assumes that the given topological circuit has been verified with respect to the 
initial circuit specification.  In general, a compressed topological circuit may bare little resemblance to the original un-optimised circuit.  Before 
a topological circuit is input to the mapping algorithm, we assume that it has been verified against the initial specification and consequently the 
qubits sets generated from the mapping algorithm will faithfully realise the original circuit when implemented in hardware.  Optimisation of 
topological circuits may employ a technique known as bridge compression \cite{FD12} which connects closed loop topological structures together.  
The algorithm detailed in this work is not currently designed to handle circuits containing bridged components, however modifying the algorithm 
to process arbitrary structures will be possible as arbitrary geometric shapes can still be decomposed into simple cycle-graphs.  These modifications 
are the subject of future work. 

\section*{Discussion}
\label{sec:conc}
In this paper we presented the first mapping algorithm that translates a topological quantum circuit into the specific measurement patterns 
required by the quantum hardware. We have also illustrated the necessary algorithm for the construction of correlation surface needed to appropriately 
track information as it propagates through the circuit.  The mapping of a topological circuit specification 
to the physical measurements in the lattice is one of many classical components necessary for realisation of quantum circuits.  This classical library consists of both office and online components.  Offline software processing consists of the compilation and optimisation of a topological quantum circuit \cite{DSMN12,PF12,FD12} and the mapping of this structure to the physical lattice.  Online classical processing 
relates to the real-time decoding and processing of error correction data using algorithms such as 
minimum weight perfect matching \cite{F13} and renormalisation group techniques \cite{GP10}.  Ultimately, 
all classical processing will need to be integrated together to form a complete software package 
for controlling topological quantum computers.

This work is not only a necessity for the future operation of topologically protected quantum computation, but it is also a requirement for 
appropriate verification packages for topological circuit synthesis.  The accurate construction of correlation surfaces allow us to compare the 
operation of a topological circuit against the abstract quantum circuit it is derived from.  This is important as optimisation techniques for topological 
quantum circuits often result in structures that bare little resemblance to the original circuit specification.  The further development of verification 
protocols is the focus of future work.


\section*{Methods}

\textbf{The Mapping Procedure.} The algorithms to compute the classical control information are presented in detail in
this section. For this purpose, we first describe several functions to translate
information between the geometric description, the cycle-graphs and the tuple
representation of logical qubits.

Function $type:V\rightarrow \{init, measure, inject, defect\}$ returns the type of a
geometric segment that was translated into a graph edge. The segment type is
used after physical qubits associated with the segment have been calculated in
order to decide which set from $q_\sigma$ they belong to.

Function $set$ takes $q_\sigma$ and the segment type as inputs and returns the
corresponding coordinate set of physical qubits that belongs to segments of $q_\sigma$ with the
specified type. For example, physical qubits from a segment of type $inject$ it will be added to the set $J^l_\sigma$, and for $init$ segments the $I^l_\sigma$ is returned.

Function $dir:V \rightarrow \{\pm w, \pm h, \pm t\}$ yields the lattice-direction
of the segment represented by the given edge. A further function $ngh^n(k),
k \in K, n \in \mathbb{Z}$ is used to indicate the $n$-th neighbour of vertex $k$
in the direction of the Hamiltonian path.

Function $coord:K\rightarrow CEL$ returns the vector representing the lattice
coordinates of a graph vertex $k \in K$. The inverse function $coord^{-1}$
is used to infer the vertex given the lattice coordinates. 

The function $mirr:K \rightarrow K$; $mirr(a)=coord^{-1}(coord(ngh^{-1}(a))+coord(ngh(a))-coord(a))$ returns vertex $a$ mirrored at the 
line through its predecessor and successor in the cycle-graph, while the $clst:K^3\rightarrow K; clst(a,b,c)$ returns either $b$ or $c$ depending which is closer to vertex $a$.



The mapping starts by constructing for each geometrical description of a logical
qubit $\sigma$ from $\mathcal{G}$ the cycle-graph $G_\sigma \in \mathcal{Y}$.
Then, the corresponding tuple $q_\sigma \in \mathcal{Q}$ is initialised by setting
$l$ to primal or dual according to the desired type and $D^l_\sigma = I^l_\sigma
= O^l_\sigma = J^l_\sigma = X^l_\sigma = Z^l_\sigma = \emptyset)$.


Sets $D^l_\sigma, I^l_\sigma, O^l_\sigma, J^l_\sigma, tube(q_\sigma)$ are
constructed by the algorithm shown in Eqn. \ref{alg:tubes}. Recall that $tube(q_\sigma)$ is either
$X^p_\sigma$ or $Z^d_\sigma$, depending on whether $q_\sigma$ is primal
($l = p$) or dual ($l = d$). The algorithm traverses the cycle-graph $G_\sigma$.
The lattice-direction $d$ associated with each edge of the graph is computed.
For each cell $cc$ of a segment, $F_{cc}$ is its complete set of physical face
qubits (see Section \ref{sec:background}). Two face qubits along the segment
are defect-internal and added to $D_{cc}$. The remaining four qubits are part
of the tube correlation surface and added to set $T_{cc}$ that is then added
to $tube(q_\sigma)$. Coordinates of injection points are always found at the
middle of the $(b,e)$ segment (Line 15). The coordinates of the qubits on $init/measure$ segments are added to the corresponding sets $I_\sigma$ and $O_\sigma$ as given by $set(q_\sigma, type)$ at Line 12;

\begin{eqnarray}
\begin{aligned}
&\textbf{Require: } q_\sigma=(l, D, I, O, J, X, Z); q_\sigma \in \mathcal{Q}; \sigma \in \mathcal{G}\\
&\textbf{Require: } G_\sigma=(K_\sigma, V_\sigma), G_\sigma \in \mathcal{Y}\\
&\text{1: } start \gets \text{ random }k \in K_\sigma\\
&\text{2: } ck \gets start\\
&\text{3: } \textbf{repeat}\\
&\text{4: } \quad d \gets dir((ck, ngh(ck)))\\
&\text{5: } \quad b \gets coord(ck); e \gets coord(ngh(ck));\\
&\text{6: } \quad \textbf{for all } cc \in (b,e) \text{ along } d\\
&\text{7: } \quad \quad D_{cc} \gets \{p | p = cc \pm 1 \text{ along } d, p \in F_{cc}\}\\
&\text{8: } \quad \quad T_{cc} \gets F_{cc} \setminus D_{cc} \\
&\text{9: } \quad \quad \textbf{if }type((ck, ngh(ck))) = defect \\
&\text{10: } \quad \quad \quad tube(q_\sigma) \gets tube(q_\sigma) \cup T_{cc} \\
&\text{11: } \quad \quad \textbf{end if} \\
&\text{12: } \quad \quad set(q_\sigma, type) \gets set(q_\sigma, type) \cup D_{cc} \\
&\text{13: } \quad \textbf{end for}\\
&\text{14: } \quad \textbf{if }type((ck, ngh(ck))) = inject\\
&\text{15: } \quad \quad J \gets J \cup \{(b+e)/2\} \\
&\text{16: } \quad \textbf{end if}\\
&\text{17: } \quad ck \gets ngh(ck)\\
&\text{18: } \textbf{until } start = ck\\
&\text{19: } \textbf{return }q_\sigma
\end{aligned}
\label{alg:tubes}
\end{eqnarray}

%
%


The sheet surfaces of each $G_\sigma$ are found by a procedure that
iteratively reduces the graph until it consists of just two vertices. The procedure constructs a sheet by finding 
one sub-sheet in each iteration, where a sub-sheet contains 
physical qubits bounded by a rectangle in either $wh$, $wt$ or $ht$ 
plane. Each sub-sheet is specified by two points $(ss^1, ss^2); ss^i \in CEL;$
where $ss^i$ define the diagonal of the rectangle and have one equal 
component. Sub-sheets found by the procedure in different steps are 
disjoint, and the union of all found sub-sheets is the complete sheet. Figure \ref{fig:tubesheet}d) illustrates a complete sheet composed of sub-sheets, and Figure \ref{fig:tubesheet}e) shows the progressive calculation of sub-sheets (both figures will be discussed in more detail later).

\begin{figure}[t!]
\centering
\includegraphics[width=15cm]{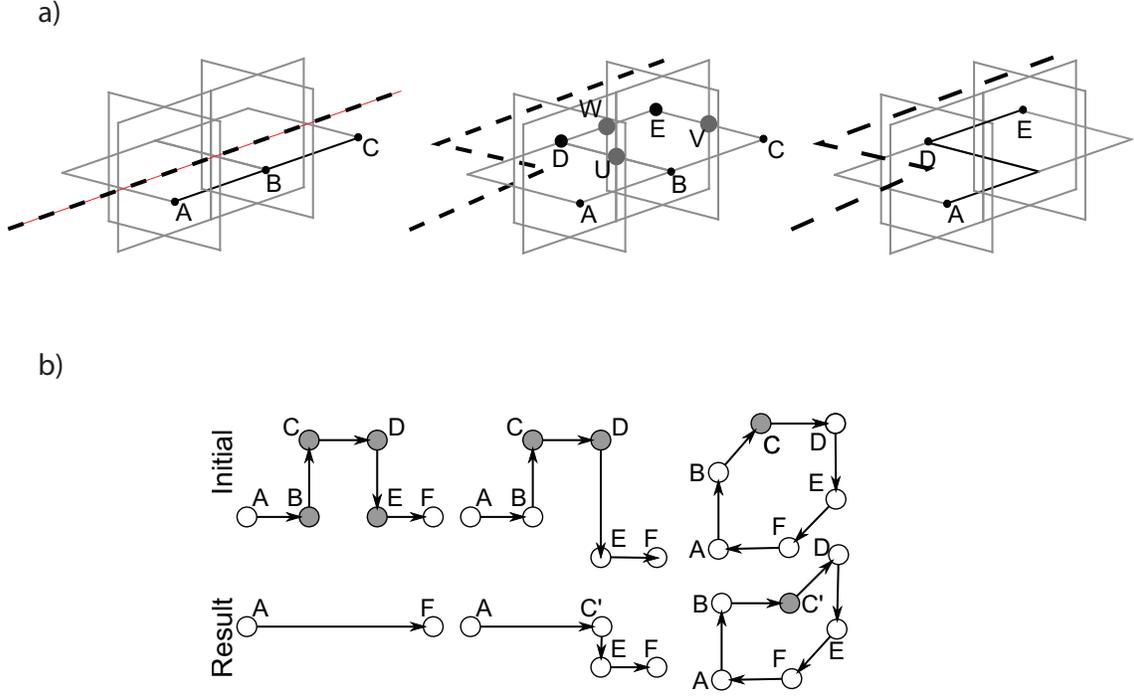}
\caption{{\bf Using defect geometries to define correlation surfaces} {\bf a)}. Changing the geometry of a defect. The trajectory of the defect (indicated by the dashed line) modifies the boundary of the sheet from the initial boundary $A,B,C$, extending the boundary $A,B,C$ involving the qubits $U,V,W$ and the resulting boundary $A,D,E$. {\bf b)} Three types of graph operations to define a surface. The first, applying $reduce(C,D)$ results in $|K_\sigma|$ being decreased by $2$, because $C,D$ are removed. After the operation, because $A,B,E,F$ correspond to co-linear lattice coordinates, $remove(B);remove(E)$ can be further applied, and $|K_\sigma|$ again decreased by $2$. In the second case, the $reduce(C,D)$ is applied. The mirrored vertices $V^{red}=\{C',D'\}$ are computed, with $coord(D')=coord(B)$, thus $V^{red}=\{C', B\}$. The vertices from $V^{red}_{ins}=\{C',B\}\setminus\{B,E\}=\{C'\}$ will be inserted, and the vertices from $V^{red}_{del}=\{B\} $will be deleted. In the last case, the effect of the $reshape(B,C,D)$ operation is that vertex $C$ is replaced by vertex $C'$. Afterwards $reduce(A,B)$ and $reduce(D,E)$ can be applied.}
\label{fig:corect}
\end{figure}

The procedure transforms the graph by eliminating or moving vertices, 
while sub-sheets are calculated. Without giving a full formal proof for 
correctness of these transformations (such a proof will require advanced 
quantum computing concepts which are out of this paper's scope), we 
illustrate the modification of a sheet boundary in Figure \ref{fig:corect}. The dashed line in the figure is indicating the direction of one defect involved in generating the sheet surface $SHEET$. By changing the direction of the defect, the boundary $A,B,C \in SHEET$ is transformed into $A,D,E \in SHEET'; SHEET' = SHEET \cup \{D,E\}$. In terms of sub-sheets, the sub-sheet $(B,E)$ is added after the change of defect direction.

The algorithm illustrated in Eqn. \ref{alg:sheet1}, used to compute the $SUBS$ set of sub-sheets for each logical qubit, will remove vertices of the graph, by constantly traversing it (Lines 3, 20), until only $2$ vertices are left (Line 2). Eqn. \ref{alg:sheet2} takes each $SUBS$ set and constructs $sheet(\sigma)$ of each logical qubit tuple $\sigma$. For the subsheet $(ss^1, ss^2)$ the function $boundingbox_{ss} =\times_{i \in \{w,h,t\}}[min(ss^1_i), max(ss^2_i)] \cap TQCC$ constructs the set of coordinates able to represent physical qubits.

Considering the structure of the lattice and the way the geometrical descriptions are constructed, it can be noticed that the cycle-graphs have an even number of vertices. It should be also noted that the $insert$ operation is not directly applied by the algorithm, and the number of vertices is modified only during the $reduce$ operation, or after the $reshape$ operations is applied.


\textit{The Reduce Operation:}
The algorithm illustrated in Eqn. \ref{alg:sheet1} modifies the graph by applying the the $reduce$ operation. The path is reduced if for $3$ consecutive edges, the first and the second edge represent opposite associated directions into the lattice (Line 10). For example, in Figure \ref{fig:corect}b) this is the case for the edges $(B,C),(C,D),(D,E)$, where $(B,C)$ and $(D,E)$ have opposite directions.

The $reduce(a,b); a,b \in K$ operation is defined as the sequential application of:
\begin{eqnarray*}
Rm=\{a,b\};\\
n_a=ngh^{-1}(a); n_b=ngh^1(b);Ng=\{n_a, n_b\}\\
V^{red}=\{clst(a, mirr(b), n_a), clst(b, mirr(a), n_b)\}\\
V^{red}_{ins} = V^{red} \setminus Ng;\\
V^{red}_{del} = Ng \cap V^{red};\\
remove(v);  \text{for all } v \in Rm\\
insert(n_a, v_i, n_b); \text{for all } v_i \in V^{red}_{ins};\\
remove(v_d);  \text{for all } v_d \in V^{red}_{del};
\end{eqnarray*}

We illustrate the reduce operation by applying it to vertices $C$ and $D$ of Figure \ref{fig:corect}b).  The sets $Rm=\{C,D\}$, $Ng=\{B,E\}$, $V^r=\{clst(C, mirr(D), B), clst(D, mirr(C), E)\}$ are constructed. Because $mirr(C)=E$ and $mirr(D)=B$, the set $V^{red}=\{B,E\}$ is equal to $Ng$ and $V^{red}_{del}=V^{red}_{ins}=\emptyset$. After the vertices from $Rm$ are removed, no further vertices are inserted or removed because the corresponding sets are empty. However, for the example in Figure \ref{fig:corect}b) this is not the case as $|V^{red}_{del}|=|V^{red}_{ins}|=1$, thus effectively removing the vertex $B$ and inserting vertex $C'$.


\textit{The Reshape Operation:}
In order for the graph to be reduced, it may be required to represent an equivalent geometrical description. Thus, vertices are not removed or deleted, but moved ($|K_\sigma|$ remains constant). The $reshape(a,b,c); a,b,c \in K$ operation is the sequential application of:
\begin{eqnarray*}
remove(b); insert(a, mirr(b), c);
\end{eqnarray*}

In the context of the algorithm in Eqn. \ref{alg:sheet1}, the function is called if during a complete graph traversal the $reduce$ operations could not be applied. For the example of Figure \ref{fig:corect}b), where $reshape(B,C,D)$ is called, the resulting graph will be obtained by removing $C$ and inserting $C'=mirr(C)$. Applying $reshape$ for a second time at the same position would undo the initial application, thus $reduce(B,C',D)$ is the inverse of $reduce(B,C,D)$. Therefore, the $start$ pivot (Line 3), used for checking if a traversal completed (Line 20), is updated (Line 24).

The maximum number of cycle traversals is $O(|K_\sigma|^2)$, when after each traversal a $reshape$ operation is required, while the number of consecutive $reshape$ operations is bounded by $|K| - 3$.

The number of vertices of the cycle-graph is initially even due to the 
shape properties of the geometrical descriptions. Moreover, it remains 
even during the execution of the algorithm. Operation $reduce$ eliminates 
exactly two vertices from the graph (set $Rm$), while the sets of further 
added and deleted vertices ($V^{red}_{ins}$ and $V^{red}_{del}$, respectively) are 
always of the same size. Operation $reshape$ does not add or delete 
vertices. However, three consecutive vertices may represent a straight line 
after a $reshape$ operation, in which case they are replaced by two 
vertices (Line 14). It can be shown that one further vertex elimination 
must follow, keeping the overall number of vertices even.


\begin{eqnarray}
\begin{aligned}
&\textbf{Require: } SUBS \\
&\textbf{Require: } q_\sigma=(l, D, I, O, X, Z); q_\sigma \in \mathcal{Q}\\
& \text{1: } sheet(q_\sigma) \gets \emptyset\\
& \text{2: } \textbf{for all } ss \in SUBS \textbf{ do}\\
& \text{3: } \quad \quad SSQ \gets \{q | coord(q) \in boundingbox(ss) \cap S^l\}\\
& \text{4: } \quad \quad sheet(q_\sigma) \gets (sheet(q_\sigma) \cup SSQ) \setminus ((sheet(q_\sigma) \cap SSQ))\\
&\text{5: } \textbf{end for}\\
&\text{6: } \textbf{return } \sigma
\end{aligned}
\label{alg:sheet2}
\end{eqnarray}


\begin{eqnarray}
\begin{aligned}
&\textbf{Require: } G_\sigma=(K, V) \text{ for } \sigma \in \mathcal{G} \\
&\text{1: }SUBS \gets \emptyset\\
& \text{2: }\textbf{while }|K| \geq 2 \text{ do}\\
& \text{3: }\quad \quad \text{start } \gets \text{ random } k \in K \\
& \text{4: }\quad \quad ck \gets start\\
& \text{5: }\quad \quad compact \gets false\\
& \text{6: }\quad \quad a \gets ngh^1(ck); b \gets ngh^2(ck)) \\
& \text{7: }\quad \quad \textbf{repeat} \\
& \text{8: }\quad \quad \quad \textbf{if } dir((ck, a)) = - dir((b, ngh(b))) \textbf{ then}\\
& \text{9: }\quad \quad \quad \quad compact \gets true\\
& \text{10: }\quad \quad \quad \quad reduce(a,b)\\
& \text{11: }\quad \quad \quad \quad SUBS \gets SUBS \cup (ngh(ck), a) \cup (ngh(ck), b)\\
& \text{12: }\quad \quad \quad \textbf{else}\\
& \text{13: }\quad \quad \quad \quad \textbf{if } dir((ck, a)) = \pm dir((a, b)) \textbf{ then}\\
& \text{14: }\quad \quad \quad \quad \quad remove(a)\\
& \text{15: }\quad \quad \quad \quad \quad compact \gets true \\
& \text{16: }\quad \quad \quad \quad \textbf{else}\\
& \text{17: }\quad \quad \quad \quad \quad ck \gets a\\
& \text{18: }\quad \quad \quad \quad \textbf{end if}\\
& \text{19: }\quad \quad \quad \textbf{end if}\\
& \text{20: }\quad \quad \textbf{until } start = ck\\
& \text{21: }\quad \quad \textbf{if }compact = false \textbf{ then}\\
& \text{22: }\quad \quad \quad reshape(start, ngh(start), ngh^2(start))\\
& \text{23: }\quad \quad \quad SUBS \gets SUBS \cup (start, ngh^2(start))\\
& \text{24: }\quad \quad \quad start \gets ngh(start) \\
& \text{25: }\quad \quad \textbf{end if}\\
& \text{26: } \textbf{end while}\\
& \text{27: }\textbf{return } SUBS
\end{aligned}
\label{alg:sheet1}
\end{eqnarray}



\section*{Acknowledgements.}
We thank A.G. Fowler for helpful discussions.  S.J.D. and K.N. acknowledge support from the HAYAO NAKAYAMA Foundation for Science \& Technology and Culture, the Quantum Cybernetics (MEXT) and FIRST projects, Japan. A.P. and I.P. acknowledge support from DFG project PO1220/3-1 and BAYFOR grant..

\section*{Author contributions.}
A.P. and S.J.D. conceived the idea.  A.P. and I.P. were responsible for algorithmic simulations.  S.J.D. and K.N. were responsible for results verification.  All authors were responsible for drafting of the manuscript. 

\section*{Additional Information.}
Correspondence and requests 
for materials should be addressed to A.P. (alexandru.paler@uni-passau.de).
\section*{Competing financial interests.}
The authors declare no competing financial interests.

\end{document}